\documentclass{mem}
\usepackage{natbib}\usepackage{txfonts}\usepackage{balance}
\usepackage{graphicx}
\idline{1}{1}
\begin{document}
\def\teff{$T\rm_{eff }$}
\def\kms{$\mathrm {km s}^{-1}$}

\title{Some caveats about the evolution of the N/O abundance and the star formation history.
}
   \subtitle{}

\author{
M. \, Moll{\' a}\inst{1} 
\and M. \, Gavil\'{a}n\inst{2}           }
\offprints{M. Moll\'{a}}
 
\institute{CIEMAT
Avda. Complutense 22
28040 Madrid, Spain
\email{mercedes.molla@ciemat.es}
\and
Universidad Aut\'{o}noma de Madrid
28049 Cantoblanco Madrid (Spain)
}

\authorrunning{Moll\'{a} \& Gavil\'{a}n }

\titlerunning{Some caveats about the N/O abundance evolution}

\abstract{
We carefully analyze how the abundance of Nitrogen over Oxygen
evolves when dependent on metallicity stellar yields with a primary
component of N proceeding from AGBs stars are used. We show
the results obtained with a chemical evolution models grid,
calculated with variable star formation efficiencies, which produce
different star formation histories. Finally we see how the N/O
abundance is related on the evolutionary history.

}
\maketitle{}

\section{Introduction}

Nitrogen abundances when compared with the Oxygen one, N/O, may, in
principle, inform about the time in which low and intermediate mass
(LIM) stars formed, and, consequently, when the widest episode of star
formation took place. This idea is based on the stellar mean-lifetimes
for LIM stars, main producers of N, which are longer than the ones of
massive stars, which eject O to the interstellar medium (ISM).

The data, however, must be carefully interpreted before to reach
misleading conclusions based in this simple scheme. If a mass of gas
form a bulk of stars simultaneously, in a Single Stellar Populations
(SSP), it is true that N will appear after O in the ISM, with a
certain time delay. However, when data proceeding from different
galaxies or different regions within a galaxy, are compared, the
previous scenario is not longer valid, since a) most of data refers to
spiral and irregular galaxies, where the star formation is a
continuous process and, due to that, there are a mix of SSPs; and b)
the observations correspond to the final stage after an evolutionary
path object, and galaxies follow different tracks. 

\section{Using the Closed Box Model}

\subsection{The basic scenario}

N may be primary, NP, created directly from the original H by the
corresponding nuclear reactions, or secondary, NS, if there exists a
seed of C or O in the gas with which the star formed. Then the
production is proportional to the original oxygen abundance.  N is
mainly produced as secondary in both massive and LIMs stars. But in
LIM stars a fraction of N is created as primary in the third dredge-up
and Hot Bottom Burning processes \cite{rv81}.  In massive stars N the
stellar rotation provokes the increase of NP, specially at the lowest
metallicities, \citep{eks08}.

\begin{center}
\begin{figure}[t!]\resizebox{\hsize}{!}
{\includegraphics[angle=-90]{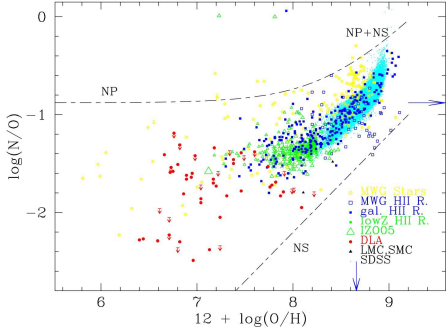}}
\caption{\footnotesize{The nitrogen over oxygen ratio, as
$\log{(N/O)}$, as a function of the oxygen abundance, as
$12+\log{log(O/H)}$, for a wide sample of data as coded: MWG stars and
HII regions; Other galaxies HII regions, (low metallicity objects
shown separately), Damped Lyman Alpha, and SDSS estimates from
\cite{lia06} }}
\label{data}
\end{figure}
\end{center}

Following the well known Closed Box Model \citep[CBM][]{tin80}, the
abundance of metals, Z, in a region may be expressed by the equation:
$Z=p ln\mu^{-1}$, where $\mu=M_{gas}/M_{tot}$ is the gas fraction in
that region and $p$ is the integrated stellar yield, the elements
newly created by a stellar generations of stars (by assuming that all
stars that produced metals are already died). By applying this
equation for N and O, and assuming that both are primary:

\begin{equation}
\frac{Z_{N}}{Z_{O}}=\frac{p_{N}}{p_{O}}= constant
\label{cbm}
\end{equation}

where $Z_{O}$ and $Z_{N}$ are the Oxygen and Nitrogen abundances,
and $p_{O}$ and $p_{N}$ are the corresponding integrated stellar yields.

If N is secondary, then:
\begin{equation}
\frac{Z_{N}}{Z_{O}}=\frac{p_{N}(O)}{p_{O}}\sim O 
\end{equation}

We plot these theoretical lines in the diagram N/O {\sl vs} O/H,
Fig.~\ref{data}.  The first case appears as an horizontal line while
the second one would be represented by a straight line with a certain
slope\footnote{Both lines are drawn with arbitrary yields}. 
When both contributions there exist, the horizontal line begins
to increase from a certain oxygen abundance. We also plot the
observations in this Fig.~\ref{data} \citep[see references
in][]{mol06}.  It is evident from this diagram that both components of
N are necessary.

Since LIM stars produce N by both processes, such as most of authors
establish \citep[see][ thereinafter GAV06, and references
therein]{gav06}, it is immediate to try to explain the data by using
the corresponding yields of LIM and massive stars. In this work we
show how N/O evolves when a contribution of NP proceeding from LIM
stars there exists.

Some people think that low metallicity objects must be young, linking
wrongly {\sl oxygen abundance-- time}, and, therefore assume that they
have not sufficient time to eject N from LIM stars. Because of that
they search for other way to obtain NP.

However, this idea that links low oxygen abundance to short
evolutionary time is a mistake: {\sl The oxygen abundance is not a
time scale.}  It is possible to have a high oxygen abundance in a very
short time, or a very low one even in an object which exists since a
long time ago if it evolves very slowly \citep{legrand00}.

\subsection{By taking into account the stellar mean-live-timescales}

It is possible to solve the equation of the CMB for both components by
assuming that there are two classes of stars, the massive ones, $m \ge
8 M_{\odot}$, which create O and secondary N, and the intermediate
stars, $ 4\leq m \leq 8 M_{\odot}$, which eject primary N, too. The
resulting expression is given by \cite{hen00} as:

\begin{equation}
Z_{N}=\frac{p_{NS}p_{C}}{2p_{O}^{2}}Z_{O}^{2}+\frac{p_{NS}p_{C}}{6p_{O}^{2}}Z_{O}^{3}
+\frac{p_{NP}}{p_{O}}(Z_{O}-Z_{\tau})e^{\frac{Z_{\tau}}{p_{O}}}
\label{NO}
\end{equation}

where $Z_{i}$ is the abundance of each element, C, N, or O, $p_{i}$ is
the corresponding yield, and the subscript $NP$ and $NP$ refer to the
the component primary and secondary.

Following that, N begins as secondary, when massive stars die, what
means that N/O evolves over an straight line with a certain
slope. After a delay time $\tau$, in which the oxygen abundance
Z$_{O}$ reaches a value $Z_{\tau}$, the NP ejected by LIM stars
appears in the ISM. Just in that moment $Z_{N}$ increases
exponentially, due to the last term of the equation, until to arrive
to the primary N level. The behavior simulates a phase change, that
occurs exactly when the first stars eject NP, time defined by the most
massive LIM stars stellar mean-lifetimes.

\begin{figure}[b]
\resizebox{\hsize}{!}{\includegraphics[angle=-90]{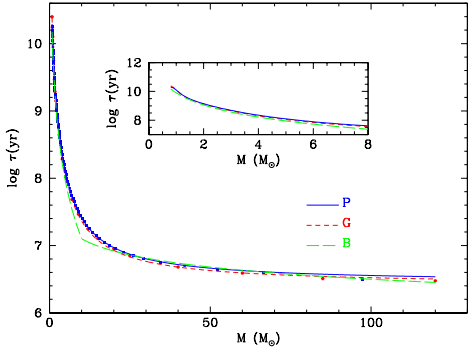}}
\caption{\footnotesize{ The stellar mean-live-timescales as a function
of the stellar mass, as obtained from Padova's group (P),
\citep{bre93}, Geneve's (G) group \citep{mey02} stellar tracks, or as
given by \cite{baz90} (B).}}
\label{tau}
\end{figure}

\begin{figure}[t!]
\resizebox{\hsize}{!}{\includegraphics[]{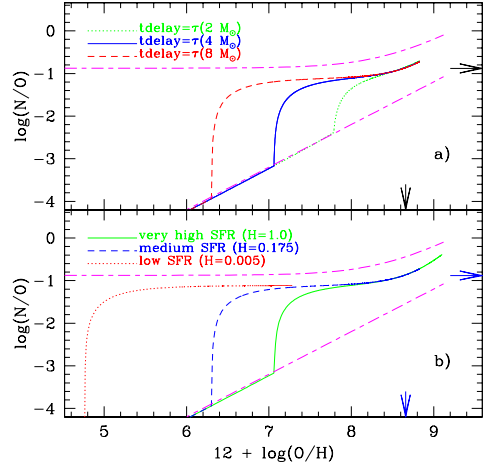}}
\caption{\footnotesize
The evolution of N/O, as log(N/O) {\sl vs} O/H, as 12+log(O/H) following
Eq.~\ref{NO} a) for stars with different masses ejecting NP, as labelled
and b) for stars of 4$M_{\odot}$ ejecting NP but where the
star formation occurs with different efficiencies.}
\label{evol_tau}
\end{figure}

If these stars have 2, 4 or 8 $M_{\odot}$, their mean-lifetimes are
$\sim 220$, 100 and 40 Myr, respectively, such as it is obtained from
the functions (Fig.~\ref{tau}) given by the usual used Padova, Geneve
stellar tracks or from \cite{baz90}. The evolutionary tracks resulting
of the above Eq.~\ref{NO} are represented in Fig.~\ref{evol_tau},
panel a).

The N/O abundance ratio increases abruptly when the oxygen abundance
is $Z_{\tau}$, which takes a different value for 2, 4, or 8
$M_{\odot}$: The smaller the stellar mass, the higher the O abundance
$Z_{\tau}$ at which the exponential function appears and the N/O
increases, such as we show in Fig.~\ref{evol_tau}a).  From this plot,
we could say that, in a Single Stellar Population in which NP is
ejected by only a value of stellar mass, if, for a given O abundance,
N/O is close to the secondary line, their stars are, in average,
younger than those ones with a higher value of N/O, nearer to the
primary line.

\subsection{The star formation efficiencies}

However, this is not the complete history.  In order to obtain the
abundance $Z_{\tau}$, following Eq.~\ref{cbm}, it is necessary to know
the gas fraction, a quantity that may change with time.  Let assume
that the galaxy starts with a value $\mu= 1$, what means that the mass
is completely in gas phase, and that it decreases when stars
form\footnote{Note that other evolutionary scenarios are possible}. If
we assume that the gas mass $g$ depends on time as: $\frac{dg}{dt}=-Hg
$, then $g=g_{0}e^{-tH}$, the mass on stars is
$s_{*}=g_{0}(1-e^{-tH})$, and $\Psi=\frac{ds}{dt}=Hg $, where H is the
efficiency to form stars. In that case the star formation law results
an exponentially decreasing function of the time:
$\Psi(t)=\Psi_{0}e^{(-t/t_{0})}$, being the time-scale t$_{0}$ the
inverse of H. If H is high, the gas will consume very quickly and,
therefore, the gas fraction $\mu=g/(g+s)$ takes a small value in a
very short time. Obviously if H is small, the gas fraction maintains
near from 1.

\begin{figure}[]
\resizebox{\hsize}{!}{\includegraphics[]{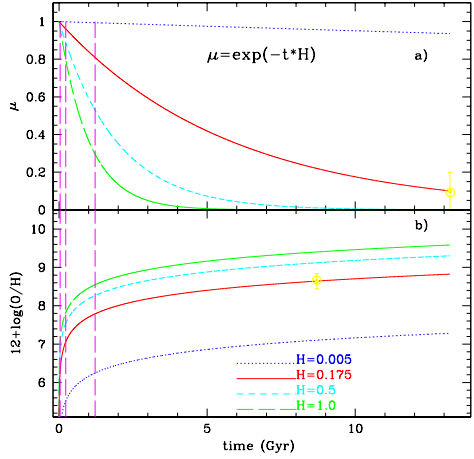}}
\caption{\footnotesize The evolution of a) the gas fraction $\mu$ and
b) the oxygen abundance $12+log(O/H)$ for different values of star
formation efficiency as labelled. The solar region value is marked as
a (yellow) solid large dot. The solar value of the oxygen abundance is
plotted at 8.7 Gyr by assuming that Sun born 4.5 Gyr ago. }
\label{mu}
\end{figure}

We show the evolution of $\mu(t)$ in in Fig.~\ref{mu}a) for some
values of H as labelled. The Oxygen abundance evolves according as
shown in Fig.~\ref{mu}b). 

Over-plotted on both graphs there are 3 lines corresponding to times,
in increasing order, equal to the mean-lifetimes of stars with 8,4 and 2
M$_{\odot}$.  We obtain 3 different values for the
oxygen abundance Z$_{\tau}$ for each evolutionary line. Equivalently,
different oxygen abundances $Z_{\tau}$ for each stellar mass are
obtained, depending on H.  If we assume that stars of 8 M$_{\odot}$
produce NP, $Z_{\tau}$ is $\sim 8$, 7.2, 6.8 or even $\leq 6$ for H$
=$ 1, 0.5, 0.172 or 0·0005 in a time of 40 Myr.  That is, the NP
appear in the ISM at an abundance that may be as lower as
$12+\log{(O/H)}=6$ if the efficiency to form stars is very low, or as
higher as 8, if the efficiency is high, but always in a same time
scale as short as 40 Myr.

We show in panel b) of Fig.~\ref{evol_tau} the evolution of the
relative abundance log(N/O) {\sl vs} the oxygen abundance
$12+log{(O/H)}$ for three values of star formation efficiencies by
assuming that NP is ejected by stars of 4 M$_{\odot}$. We see a
similar behavior to that one shown in panel a).

\section{The metallicity dependent stellar yields}

The previous section results are obtained by taking constant effective
yields $p_{NP}$, $p_{NS}$ and $p_{O}$. Actually, the stellar yields
depend on the metallicity Z with which stars form. And therefore the
effective yield for a single stellar population also depends on Z. In
the case of LIM stars, this question is even more important since,
obviously, the proportion NP/N must change with Z. If the star has a
very low Z, the abundance of O, the seed for the NS, is also low, so
most N is created as NP.

This is clearly seen in Fig.2 from GAV06 where this proportion NP/N is
represented as a function of the stellar mass for some sets of stellar
yields from the literature. In panel a) the solar abundance yields
from \citet[][hereinafter BU,VG, and MA, respectively]{gav05,vg,ma01}
are represented.  In panels b), c) and d) the same is shown for two
values of Z for each one of these sets separately, as labelled. In all
cases, NP/N is higher for the low Z set than for the solar abundance
one.

The dependence on Z of the integrated yields for a simple stellar
population is shown in Fig.3 from GAV06.  In panel a) the integrated
yield of N tends to be larger for higher Z. In panel b) the ratio
$p_{NP}/p_{N}$ clearly decreases with Z, such as it must do. The
different sets, however, have different behaviour. BU has smaller
total yields of N than VG and more similar to values from MA, but the
dependence on Z is smoother that this one shown by VG and MA.

\begin{figure}[t!]
\resizebox{\hsize}{!}{\includegraphics[]{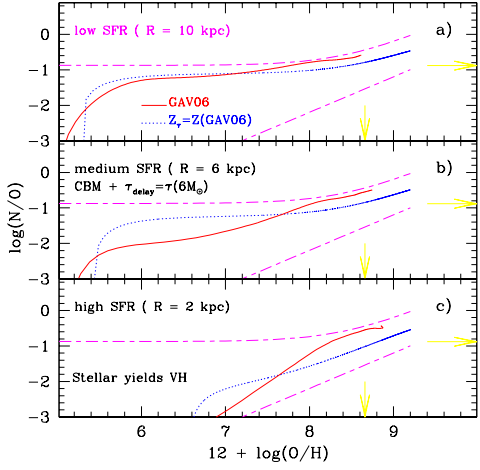}}
\caption{\footnotesize{The evolution of log(N/O) as a function of
12+log(O/H) for different star formation rates: a) low,
b) intermediate and c) high efficiency, respectively. Dotted lines
are the CBM results while the solid lines are the results obtained by
GAV06 for different regions of the MWG disk.}}
\label{Zdep}
\end{figure}

By using dependent yields on the CBM equation of the previous section
the results shown in Fig~\ref{Zdep} are obtained. Results for
different efficiencies H, low, solar and high, are in panels a), b)
and c) as dotted lines.  The main effect of changing yields with Z is
that evolutionary tracks elongate when the star formation rate is
strong, as it occurs in c).  Therefore the efficiency to form stars,
which define the shape of the star formation history, the level of
starburst with which the stars form, is essential to obtain an
elongated or flat evolutionary track in the plane NO-OH.

\section{The chemical evolution model grid}
\subsection{The local universe}

Until now only CBM results have been analyzed. However modern chemical
evolution models are usually numerical codes that include a larger
quantity of information, taking into account the mean-lifetimes of
stars, the IMF, the stellar yields, a large number of elements, etc.
We have over-plotted in Fig.~\ref{Zdep}, with solid lines, the results
from GAV06, obtained by using the multiphase chemical evolution model
\citep{fer94,mol05} with the VG stellar yields for different regions
of the Milky Way Galaxy (MWG), located at 10, 6 and 2 kpc of
galactocentric.  They are similar to those for the CBM.

\begin{figure}[t!]
\resizebox{\hsize}{!}{\includegraphics[angle=-90]{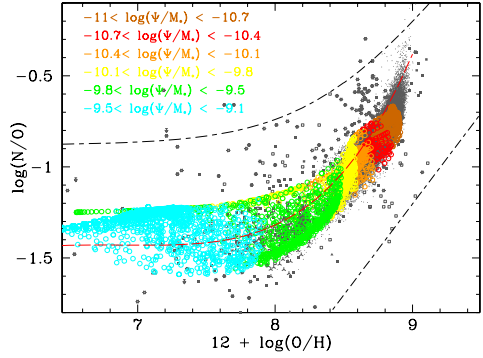}}
\caption{\footnotesize{Multiphase chemical evolution model results for
the present time for a grid of 440 galaxies as open dots. Small dots 
in the background are the SDSS data and the other symbols are extragalactic
and Galactic HII regions data as labelled.}}
\label{t13}
\end{figure}

Following the same scenario than GAV06, we have compute models 
for 44 sizes or masses of galaxies and 10
values of efficiencies in the range [0,1] for each one of them
\citep[see][for details]{mol05,mol06}.  The results for the present
time are shown in Fig.~\ref{t13} with large open dot while the small
(grey) dots represent the SDSS data and other HII regions data from
the literature. Our results even reproduce the high dispersion
observed in the plane NO-OH: there are two regions where the
dispersion is higher, one located around (7.8, -1.45) and a second one
around (8.7,-1), just where the low mass and the bright massive
galaxies fall.

These models demonstrate clearly that efficiencies to form stars have an
important role in the evolution of the tracks in the plane NO-OH, and are
essential to reproduce the data.

\subsection{The star formation history role}

\cite{mal07} have shown that the location of a galaxy in the plane NO-OH
is related with the specific star formation rate, the star
formation rate by stellar mass unit, sSFR. They found that the highest
values of N/O correspond to the smallest values of sSFR.
\begin{figure}[]
\resizebox{\hsize}{!}{\includegraphics[angle=0]{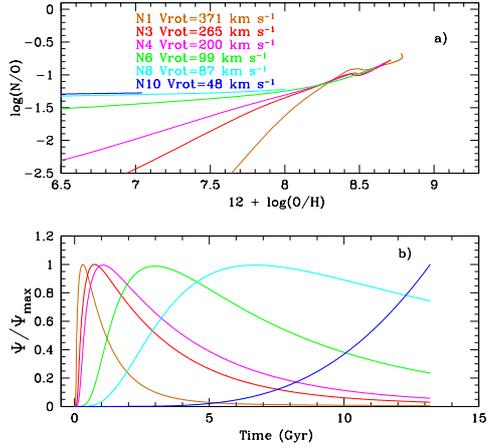}}
\caption{\footnotesize{a) Evolutionary tracks for six different theoretical
galaxies in the plane No-OH, b) Star formation histories for the same 6 
models}}
\label{no_sfr}
\end{figure}

Since we also reproduce this trend, as we may see with different code
of colors in Fig.6, we may explain the subjacent reasons for this
correlation.  In our models the smallest sSFR occur in galaxies where
the star formation was high in the past, in the earliest times of
evolution. The gas was rapidly consumed and therefore the star
formation rate decreased since then, showing now very small values. On
the opposite side, when the efficiency to form stars is very low, the
star formation history increases with time, and, because of that, the
present star formation rate is high. The evolutionary track of an
object and its final point in the NO-OH plane depends on the star
formation history suffered by the galaxy.  In order to demonstrate
that, we show in Fig.~\ref{no_sfr}a) 6 different evolutionary tracks
in the plane NO-OH and the corresponding star formation rate in panel
b) of the same figure.

The low mass galaxies with low efficiencies have not had a high star
formation in the past, and they may considered young from the point of
view of the averaged age for most of their stars, created mainly in
recent times. But they also have old stars able to eject nitrogen,
that was processed as primary, since in that moment Z was very low.

\section{Conclusions}

The present day data as well as the high-redshift trends in the
plane O/H-N/O may be reproduced with chemical evolution models
using stellar yields where LIM stars create a certain quantity of NP

Differences in the star formation histories of galaxies are
essential to reproduce the data and the observed dispersion,
the present position of a galaxy in the diagram N/O vs O/H being
determined by this evolutionary history 

It is possible to have NP ejected by LIM stars even at a low O abundance
since O abundance is not a time scale.  

\bibliographystyle{aa}

\end{document}